%% file: main.tex
\documentclass[preprint,journal]{vgtc}       

\ifpdf
  \pdfoutput=1\relax                   
  \usepackage{graphicx}                
  \DeclareGraphicsExtensions{.pdf,.png,.jpg,.jpeg} 
\else
  \usepackage{graphicx}                
  \DeclareGraphicsExtensions{.eps}     
\fi%

\graphicspath{{figures/}{pictures/}{images/}{./}} 

\usepackage{enumitem}
\usepackage{amsmath}
\usepackage{microtype}                 
\PassOptionsToPackage{warn}{textcomp}  
\usepackage{textcomp}                  
\usepackage{mathptmx}                  
\usepackage{times}                     
\usepackage{cite}                      
\usepackage{tabu}                      
\usepackage{booktabs}                  
\usepackage{subfig}                    

\onlineid{1277}

\vgtccategory{Research}
\vgtcpapertype{algorithm/technique}

\usepackage{algorithm}
\usepackage[noend]{algpseudocode}
\algnewcommand\algorithmicforeach{\textbf{for each}}
\algdef{S}[FOR]{ForEach}[1]{\algorithmicforeach\ #1\ \algorithmicdo}

\newcommand{\anovatest}[4]{\textit{F(#1)} = #2; $p<$#3; $\eta^2_p$=#4}
\newcommand{\ARTanovatest}[3]{\textit{F(#1)} = #2; $p<$#3}
\newcommand{\msd}[2]{\textit{Avg}=#1; \textit{SD}=#2}
\newcommand{\msdp}[3]{\textit{Avg}=#1; \textit{SD}=#2; $p<$#3}
\usepackage{xfrac}
\usepackage{macros}
\usepackage{comment}

\title{Timeline Design Space for \\ Immersive Exploration of Time-Varying Spatial 3D Data}

\author{Gwendal Fouch\'e, Ferran Argelaguet, Emmanuel Faure and Charles Kervrann}
\authorfooter{
\item
 Gwendal Fouch\'e and Ferran Argelaguet are with Inria, Univ Rennes, IRISA, CNRS, Rennes, France. 
\item
 Charles Kervrann is with Inria Centre Rennes-Bretagne Atlantique, UMR144 CNRS Institut Curie, PSL Research University, Sorbonne Universités, Paris, France.\\
 E-mail: name.surname@inria.fr.
\item
 Emmanuel Faure is with LIRMM,CNRS,Montpellier, France. E-mail: emmanuel.faure@lirmm.fr.
}
\abstract{
Timelines are common visualizations to represent and manipulate temporal data, from historical events storytelling to animation authoring.
However, timeline visualizations rarely consider spatio-temporal 3D data (e.g. mesh or volumetric models) directly, which are typically explored using 3D visualizers only displaying one time-step at a time.
In this paper, leveraging the increased workspace and 3D interaction capabilities of virtual reality (VR), we propose to use timelines for the visualization of 3D temporal data to support exploration and analysis.
First, we propose a timeline design space for 3D temporal data extending the timeline design space proposed by Brehmer et al.~\cite{brehmer_timelines_2017}.
The proposed design space adapts the scale, layout and representation dimensions to account for the depth dimension and how the 3D temporal data can be partitioned and structured. In our approach, an additional dimension is introduced, the \textit{support}, which further characterizes the 3D dimension of the visualization.
To complement the design space and the interaction capabilities of VR systems, we discuss the interaction methods required for the efficient visualization of 3D timelines. 
Then, to evaluate the benefits of 3D timelines, we conducted a formal evaluation (n=21) with two main objectives: comparing the proposed visualization with a traditional visualization method (using a temporal slider); exploring how users interact with different 3D timeline designs. Four tasks, relevant with temporal data, were considered: locating a time point, counting occurrences, finding temporal patterns and finding a global maximum.
Our results showed that time-related tasks can be achieved more comfortably using timelines, and more efficiently for specific tasks requiring the analysis of the surrounding temporal context.
Though the comparison between the different timeline designs were inconclusive, participants reported a clear preference towards the timeline design that did not occupy the vertical space.
Finally, we illustrate the use of the 3D timelines to a real use-case on the analysis of biological 3D temporal datasets in which domain experts in cell imaging were involved in the design and evaluation process.
} 


\usepackage{multirow}
\teaser{
  
  \centering
  \includegraphics[width=\linewidth]{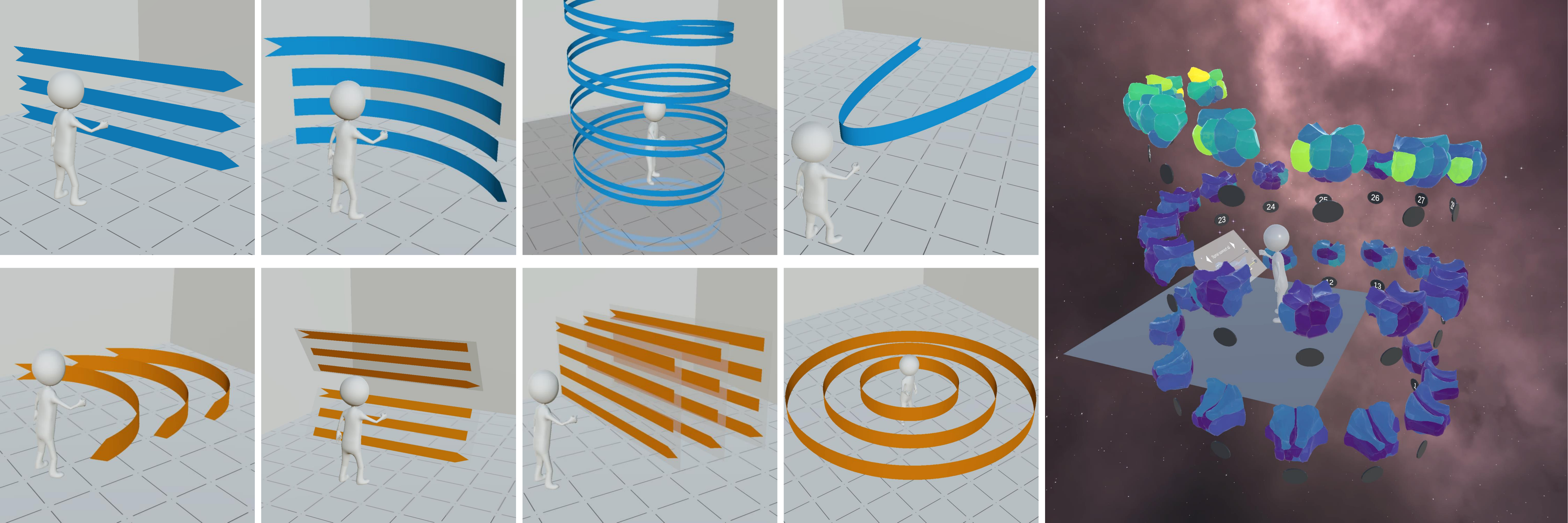}
     \caption{Illustration of the proposed 3D timeline design space for immersive exploration of time-varying spatial 3D data. The first row showcases different timeline representations, i.e. guiding curve of the timeline; while the second row showcases different support configurations. The rightmost image ilustrates the use of an helicoid representation for the exploration of an time-varying embryo imaging dataset. The viridis colormap is used to encode the elongation ratio of cells involved in morphogenetic movements.}
   \label{fig:teaser}

}

\vgtcinsertpkg

\begin{document}



\maketitle

\input{chapters.tex}



\bibliographystyle{abbrv-doi}

\bibliography{biblio}



\end{document}

%% file: chapters.tex
\section{Introduction}

In the biology and medical domain, 3D imaging breakthroughs have notably increased the availability of time-varying spatial 3D data creating roadblocks and new challenges for researchers in image analysis, and visualization in order to fully exploit the large amount of data being produced~\cite{kriston2017getting, li20113d}. However, while automatic and semi-automatic methods can be used to analyze such datasets, the direct exploration by domain experts remains a necessary step during the analysis process.
The most common approaches for the visualization of time-varying spatial 3D data, remain the use of temporal sliders~\cite{duran_visualization_2019}, animation~\cite{johnson_bento_2019} or juxtaposition of time points~\cite{lu2008interactive}. However, such method presents a number of limitations, such as relying on the short term memory to assess the temporal changes, and the limit of time points that can be juxtaposed due to the limitations of display size. One potential solution is to consider timeline visualizations to represent the evolution of the temporal dimension, however, while timelines are powerful visualizations for temporal events, they are less adapted for 3D spatial data. 


This paper proposes a extended timeline design space for immersive 3D temporal data, inspired from the design space described by Brehmer et al.~\cite{brehmer_timelines_2017}. Our design space extends the original design space, adapting them to additional workspace dimension, defining how data can be partitioned and structured in a 3D environment.
Moreover, due to the specific characteristics of interaction of immersive contexts and in order to take full advantage of 3D timelines, we discuss the basic set of interaction techniques to explore and manipulate them, relying on methods of the field of Immersive Analytics~\cite{fonnet_survey_2019}, and propose a set of design criteria to drive the design process.

In order to explore the potential benefits of 3D timelines, we conducted two evaluations. The first evaluation, was a formal summative evaluation, which explored how 3D timelines could improve data exploration tasks. For this evaluation, mainly VR experts were considered, and we mainly focused on performance and usability.
The second evaluation, was a qualitative evaluation conducted in collaboration with experts in biology. The objective was to gather feedback on how 3D timelines could be included in their analysis workflow, precisely, on how they could take advantage of 3D timelines to ease the exploration and validation process on their 3D temporal imaging data.

\section{Related Work}\label{relatedWorks}
\subsection{Visualization of High Dimensional Data}

High dimensional data poses a challenging visualization problem not only due to the raw data size but also due to the heterogeneity of the different dimensions constituting these datasets.
%
%
These dimensions are spatially and temporally referenced, represent numerical or categorical information, and come from various sources, such as imaging, simulations or annotation processes.
%
%
%
Andrienko et al.\cite{andrienko_space_2010} presented various approaches to handle this challenge: data aggregation and more generally dimensionality reduction; semi or fully automatic feature extraction; and juxtaposition.

First, dimensionality reduction approaches focus on extracting and displaying the most relevant information with a lower-dimensional representation in order to make it easier to apprehend or to help in classification tasks. 
For example, Dem\v{s}ar and Virrantaus~\cite{demvsar2010space} visualize ocean vessels trajectory sets by aggregating and summarizing the data as a volume by computing the density of movement.
The result is thus less cluttered and more legible.
Woodring and Shen~\cite{woodring_multi-variate_2006} proposed a general visualization scheme allowing the projection of several variables, combined with different volume operations into a 3D volume.
Dimensionality reduction can also rely on linear transformations.
Principal component analysis is a classic example in the analysis of multidimensional points clouds. 
Derived algorithms~\cite{comon_independent_1992, ivosev_dimensionality_2008} further improved the method, better conserving the variance of the original data.
Other methods use non-linear transformations to reduce dimensions, such as methods based on Fisher Linear Discriminant Analysis~\cite{mika_fisher_1999} or the locally linear embedding~\cite{roweis_nonlinear_2000}.
The second approach consists in extracting specific features or patterns from the data.
This can result in abstract representations that may complete the analysis on a direct depiction of the data.
For example, TransGraph~\cite{gu_transgraph:_2011} extracts transitions in features or data items in time-varying volumetric data to produce a graph juxtaposed with the 3D rendering. 
Finally, juxtaposition helps representing and exploring high dimensional data by displaying different views and projection of the data separately.
Regarding this topic, Munzner~\cite{munzner_visualization_2014} proposed four design choices for coordinated juxtaposed views.
The different models were characterized by whether the encoding used is different or not, and whether the data displayed in each view is complete, a sub-set or a partition.
Respectively in accordance with the rules of Diversity and Consistency proposed by Baldonado et al.~\cite{wang_baldonado_guidelines_2000}, juxtaposing and coordinating several dimensionality-reduced views is a reliable option to visualize high dimensional data.

\subsection{Visualization of Time-Varying Spatial 3D Data}

Kim et al.~\cite{kim_comparison_2017} define spatial 3D data as ``data with an inherent, meaningful width, height, and depth, where the relative position of things, their length, surface size and shape, etc. all might matter to a user'', and extend this definition to time-varying spatial 3D data, shortened respectively as S3D and S4D.
S3D data considers four main categories: 3D paths (e.g. curves, trajectories), glyph or point clouds, surface data and volumetric data~\cite{kim_comparison_2017}.
Moreover, categorical or numerical data are often present in such datasets, added through automatic or manual annotation processes, which increase the complexity of the data and the representation.
The visualization of S4D data is challenging, as their large size, high resolution and high density imply not only potential technical difficulties with rendering, but also issues with level-of-detail~\cite{luebke2003level} and occlusion~\cite{elmqvist2008taxonomy}, that can disturb the exploration of the data. 

For example, Duran et al.~\cite{duran_visualization_2019} proposed a tool to visualize 3D temporal datasets of molecule simulations, juxtaposing a 3D data render with charts to visualize one-dimensional numerical information.
%
%
The small multiple approach pushes this method to the extreme by coordinating a partition of the data.
Bento Box~\cite{johnson_bento_2019} juxtaposed multiple instances of a same volumetric dataset under different parameters.
Relying on an immersive environment and 3D user interface, the different views are displayed as a grid, interactions on time and rotation are coordinated.

The visualization of S4D data typically considers the visualization individual time-steps, yet, the visualization of multiple time-steps could enable a better identification of temporal events, thus a 3D timeline could be directly use. 

\subsection{Timeline}\label{refTimeline}


Timelines are a classic visualization for temporal data, which are mainly used to represent series of events linearly or in form of a tree.
Timelines have a wide range of design choices~\cite{brehmer_timelines_2017}, and this range allows to create expressive representations.
Hence, they are often used for summarizing events, storytelling or historical summaries, but also for planning, using calendars or Gantt charts for instance.
Narrative visualizations are also often used to present data and information in an attractive yet understandable way, and timelines can be designed to balance perceptual and narrative effectiveness for this purpose.
For example, TimelineJS~\cite{TimelineJS} and TimelineSetter~\cite{TimelineSetter} are tools used in general media to generate slideshow timelines to describe long narratives, gathering numerous major events.
In a context of data analysis, Lifeflow~\cite{wongsuphasawat_lifeflow_2011} and TimelineTrees~\cite{burch_timeline_2008} aggregate data using tree structures, and use timelines to explore the temporal aspect of these data.

Brehmer et al.~\cite{brehmer_timelines_2017} proposed a classification of timelines based on a review of 263 timeline designs, proposing a 3-dimension design space.
The first dimension is the \textit{Representation} of the timeline, i.e. its ``most visually salient aspect, its guiding visual metaphor''. 
The most common and versatile representation of a timeline is linear, yet radial or spiral representations can be more adapted to show periodic data.
Other representations can rely on grids, as for calendars, or on arbitrary curves designed to support a narrative.
The second dimension is the \textit{Scale}, which determines the link between temporal and displayed distance; it is characterized notably by its reference point and the function mapping the temporal dimension (e.g. linear or logarithmic). 
The third dimension is the \textit{Layout}, which describes if the display shows one timeline or multiple faceted ones, if the timelines are segmented or not.
This design space includes some non-viable designs, yet describes comprehensively most timeline designs. 
Nonetheless, 3D timeline designs are not approached in this design space, and are actually lightly explored in the literature.
The few examples include Beedocs~\cite{Beedocs}, a timeline authoring tool using 3D animations to explore a 2D timeline, HeloVis~\cite{cantu2018helovis} which displays radar signal data on an helical visualization or works by Kullberg et al.~\cite{kullberg1995dynamic}, who use the horizontal space of a 3D environment to display time.

To help the exploration of timelines, interaction methods adapted to the visualization have been developed.
The most common interaction is based on scrolling or sliders to move in time.
For example, works by Charles et al.~\cite{charles_timeline-based_2011} or Card et al.~\cite{card_time_2006} use timelines to give an overview of events while using sliders to explore data in time, that is displayed in a juxtaposed view.
TimeZoom~\cite{dachselt_timezoom_2006} also relies on horizontal scrolling to move in time, yet also base several other interactions on \textit{regions of focus}. 
Such interactions include scrolling in time by dragging the region, zooming on the region or the whole timeline in order to variate levels of detail, and finally means of creation, deletion and edition of such regions.
The need of enhanced interactivity led to the use of various user interfaces for timelines.
As such, Morawa et al.~\cite{morawa_combining_2014} proposed Time Beads to interact with the timeline using a touch user interface to support intuitive manipulation of time points, and Drossis et al.~\cite{drossis20133d} proposed to explore the temporal information through a \textit{time-tunnel}, displayed in an immersive 3D environment.

However, S4D data is rarely considered on timeline visualizations.
A few examples can be found in contexts of 3D model editors, as proposed by Denning et al.~\cite{denning_meshflow_2011} or Dobo\v{s} et al.~\cite{dobos_3d_2014}, that use timelines to display the step-by-step construction of the model. 
Although, S4D visualization tools sometimes use timelines as sliders, juxtaposing 1-dimension additional information with the 3D spatial data, as in Duran et al.~\cite{duran_visualization_2019} work or RubberSlider \cite{cheymol2022rubber}, which uses a VR adjustable slider to explore a timeline and display 3D data.
The lack of design guidelines for the visualization of S4D data using timelines motivated the extension of the design space from Brehmer et al.~\cite{brehmer_timelines_2017}. 
Yet, the visualization of such 3D timelines would require large displays to ensure that a reasonable number of time points can be explored and visualized efficiently, thus, we leveraged the increased workspace size, depth cues~\cite{loomis_visual_2003} and navigation techniques~\cite{laviola20173d} provided by immersive virtual reality systems to provide efficient visualization methods.

\section{A Timeline in a 3D Environment}

This section introduces our design space for 3D timelines, which extends the 2D timeline design space~\cite{brehmer_timelines_2017} for 3D environments, and notably immersive environments.
The extension considers the additional spatial dimension in which data can be laid enabling the direct display of the temporal dimension of the data.

As mentioned in the introduction, we focus on S4D datasets. 
We assume that the temporal component of such dataset is discrete, or at least it can be discretized.
As such, we can construct a timeline in which every time point is populated by a \textbf{3D snapshot}, i.e. the state of the 3D temporal dataset at an instant.
%
%
We consider that some datasets are composed of a set of objects, i.e. shapes with the own semantic value, and define, assuming sufficient tracking information, \textbf{4D objects} that span along multiple time points. 
\ferran{Can be removed if needed, too specific $->$} These objects can be defined by segmentation process for imaging data~\cite{guignard_contact-dependent_2018}, or by construction in a surface-based 3D model~\cite{denning_meshflow_2011,dobos_3d_2014}.

Finally, in order to exploit the interaction capabilities provided by IVR systems and to take advantage of the 3D timelines, we detail the set of interaction tasks needed for their exploration.

\subsection{Extending the 2D Timeline Design Space}

As presented in Section~\ref{relatedWorks}, Brehmer et al.~\cite{brehmer_timelines_2017} proposed a design space for 2D timelines with the three following dimensions: \textit{scale}, \textit{layout} and \textit{representation}
In this part, we adapt the design space to 3D timelines.
The two first characteristics are mainly similar for 2D and 3D timelines, we added those parts for completeness of the design space, also adding a discussion for 3D temporal data. 
Then, we extended the design space by adapting the \textit{representation} dimension choices, yet also proposing other types of guiding curves adapted to a 3D environment.
Finally, we introduce a fourth dimension to this design space, as a result of using a 3D display, which describes the supporting plane on which timeline branches are laid: the \textit{support} dimension.

\subsubsection{Scale: Displaying Time}

The \textit{scale} dimension represents the relation between the displayed distance and the temporal distance.
Although events will be laid in 3D space, this relation will remain.
%
%
Therefore, the \textit{scale} dimension remains characterized as defined by Brehmer et al.~\cite{brehmer_timelines_2017}:

\begin{itemize}
    \item A \textbf{chronological} scale spaces time points according to the actual temporal distance, either \textbf{linearly} or \textbf{logarithmically}, depending on the distribution of the events.
    \item A \textbf{relative} scale arranges the time points according to a baseline event. It can notably be used to compare multiple timelines regarding this event. 
    \item With a \textbf{sequential} scale, the distance between time points is fixed and does not correspond to the temporal distance. A similar approach consist in encoding the duration between time points on the timeline as well.
\end{itemize}

\subsubsection{Layout: Arranging Time Points}

The layout dimension corresponds to how the timeline is partitioned in the display. 
Brehmer~\cite{brehmer_timelines_2017} describes several characteristics of a 2D timeline layout.
A first main characteristic of a layout is whether it is \textbf{segmented} or not. 
Segmentation is used to cope with spatial organization issues, such as the limitation in size of a 2D display, but also for analytic purposes, for example to separate the timeline to make some periodic events more salient.
The timeline layout can also be qualified as \textbf{unified} or \textbf{faceted}, i.e. showing a timeline with one or several lines; the latter can be useful for comparing data from different sources over time.
Whether derived from segmentation or faceting, we refer to timeline parts as \textit{branches}.

These two characteristics can be used to describe 3D timelines as well.
Similar factors are taken into account for the choice in layout.
Segmentation helps in solving issues of juxtaposition of visualization and optimization of the large workspace offered by 3D environments.
Its analytical purpose is notably relevant in case of periodic data.
On the other hand, the choice between faceted or unified layout relies on the nature of the data and use case.
A unified timeline design is more suited to focus on the dynamic features of one region of interest or group of objects, while a faceted layout allow the comparison of several ones.
Faceted layouts are also adapted to compare of multiple datasets, as in such cases, the most common task is comparison.
In any case, faceting or segmenting the layout tends to clutter the workspace.
Attention should be given to properly indicate the origin and characteristics of each facet or segment as not to overwhelm the user.

\subsubsection{Representation: Shaping the Timeline}

\begin{figure}[t!]
\centering
\includegraphics[width=0.23\textwidth ]{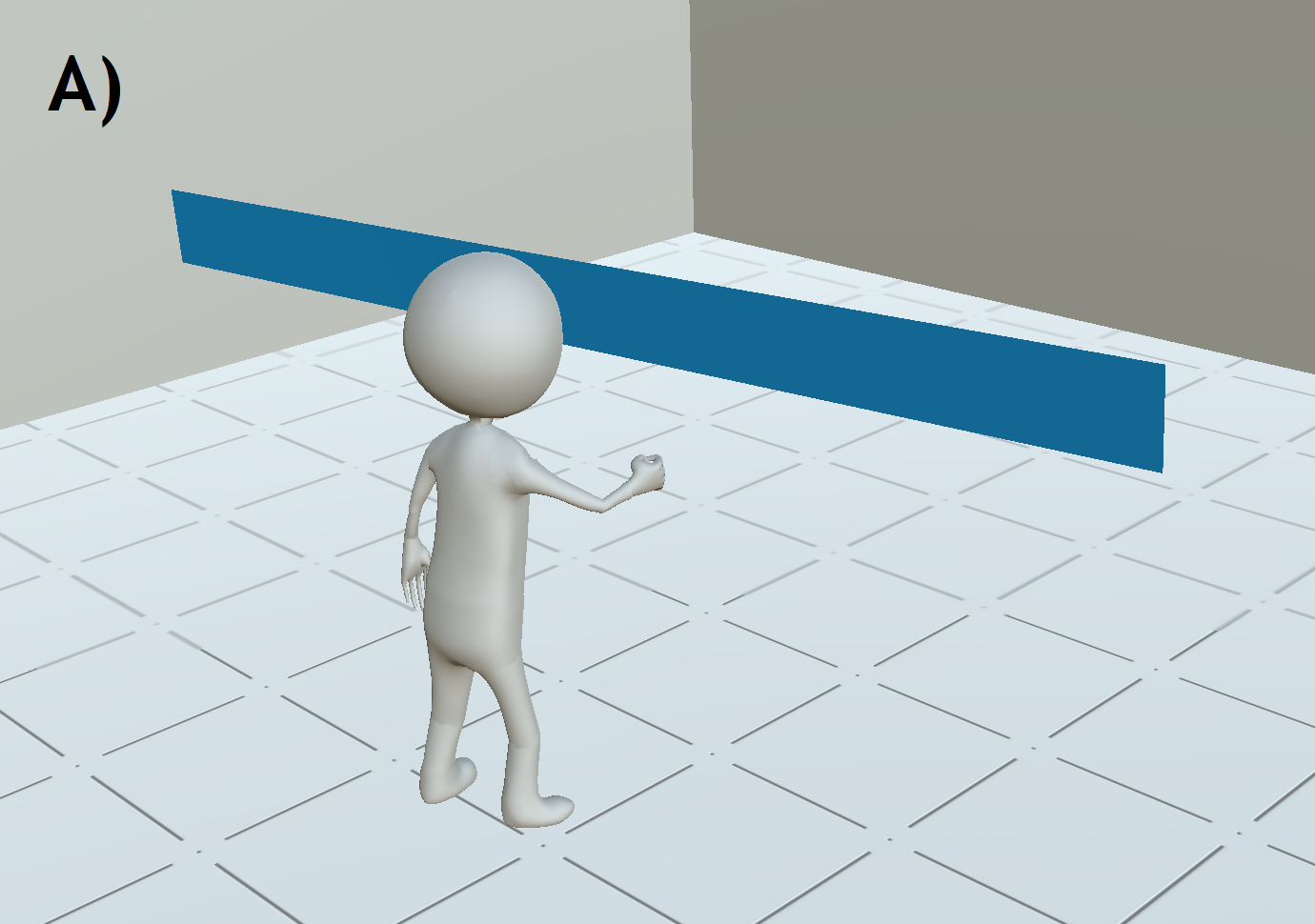}
\hspace{0.05cm}
\includegraphics[width=0.23\textwidth ]{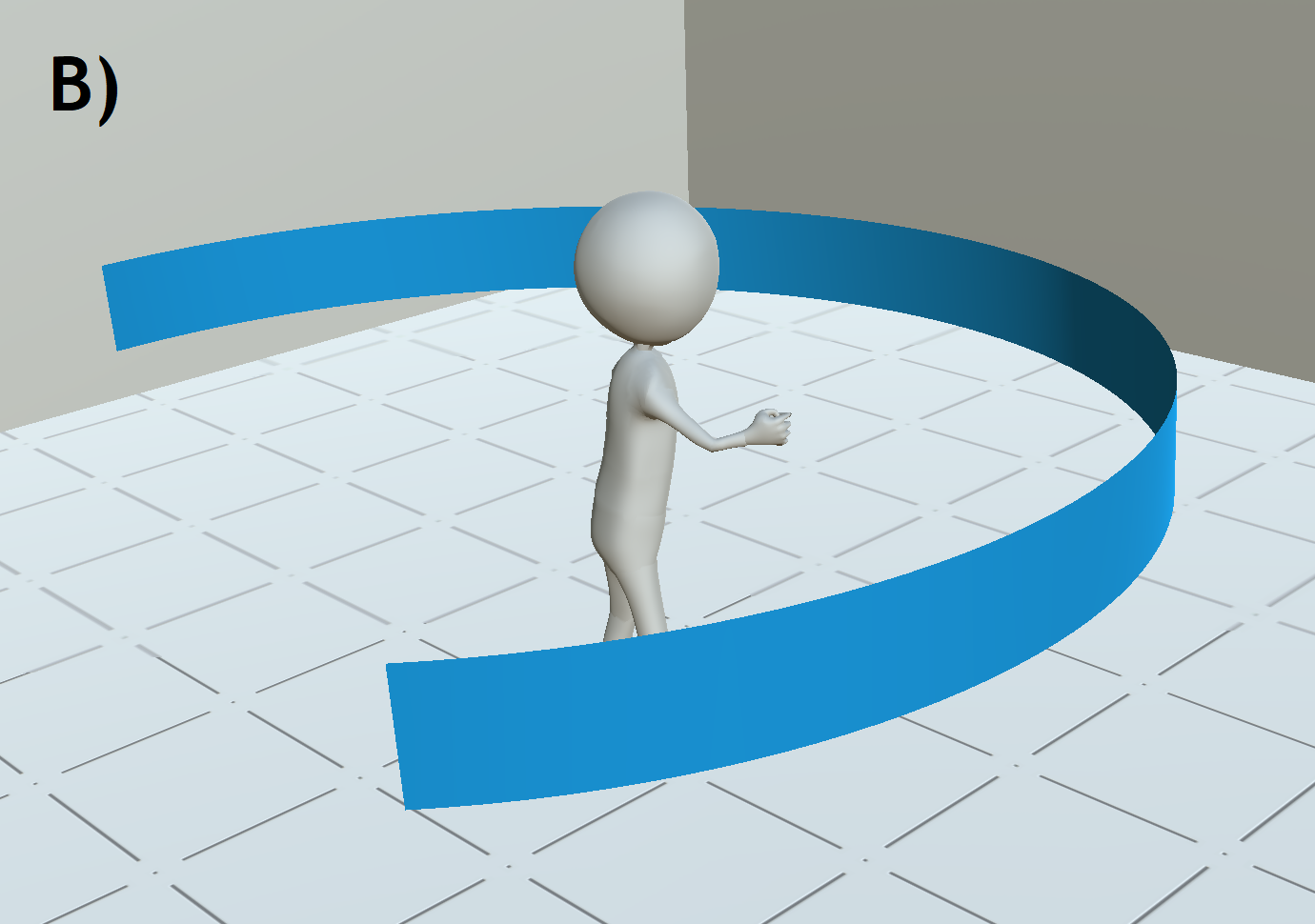} \\ \vspace{0.2cm}
\includegraphics[width=0.23\textwidth ]{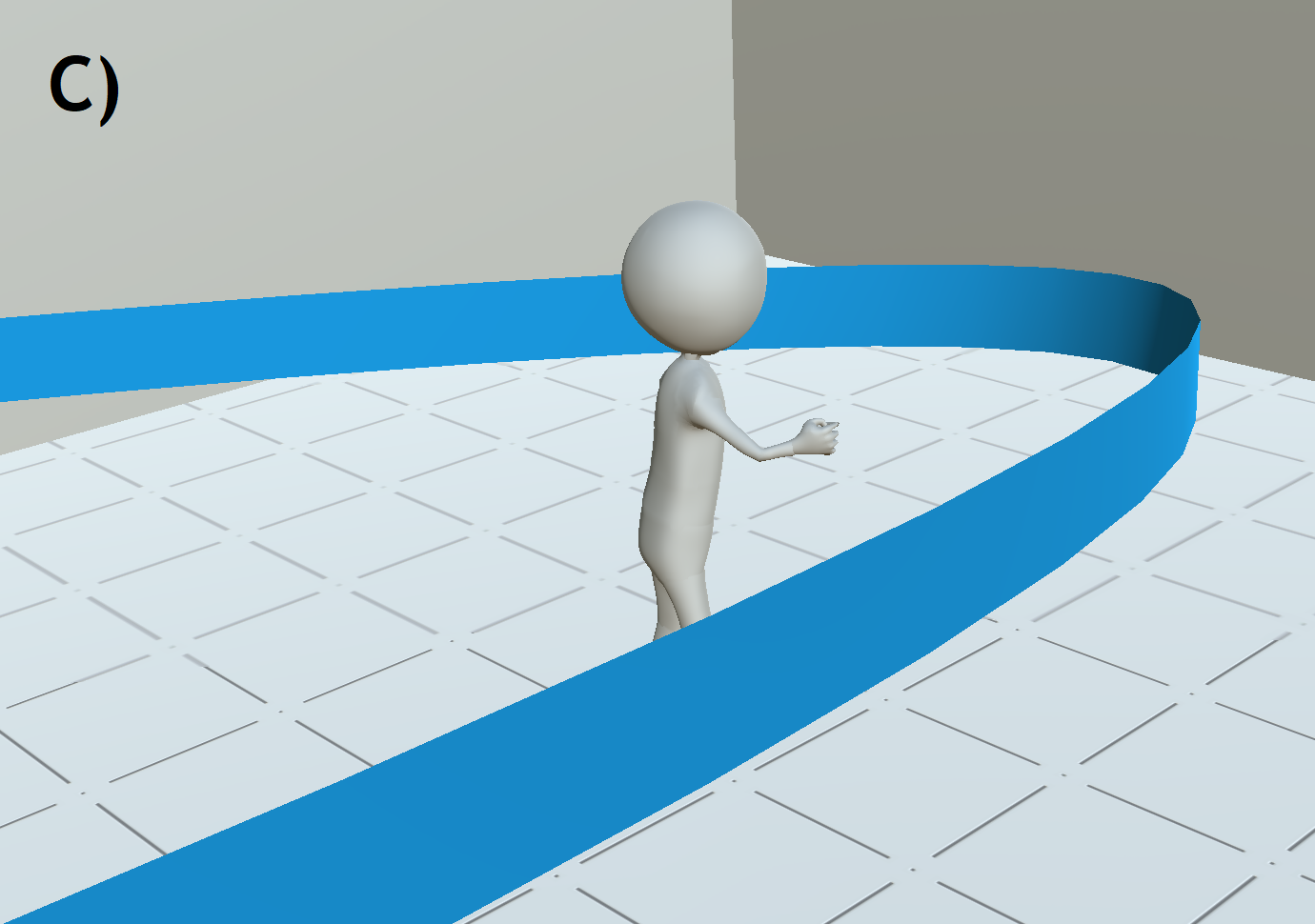} \hspace{0.05cm}
\includegraphics[width=0.23\textwidth ]{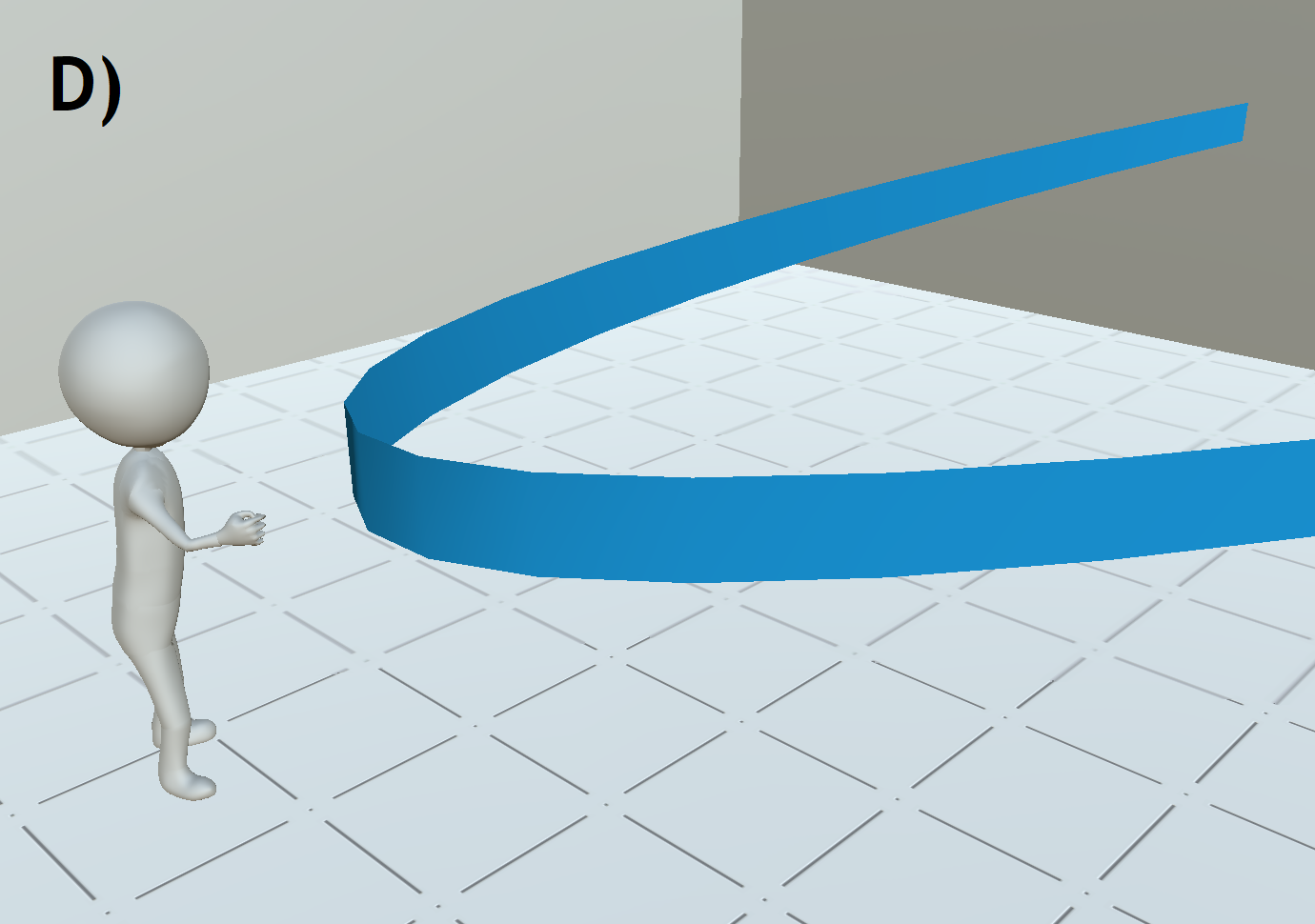} \\ \vspace{0.2cm}
\includegraphics[width=0.23\textwidth ]{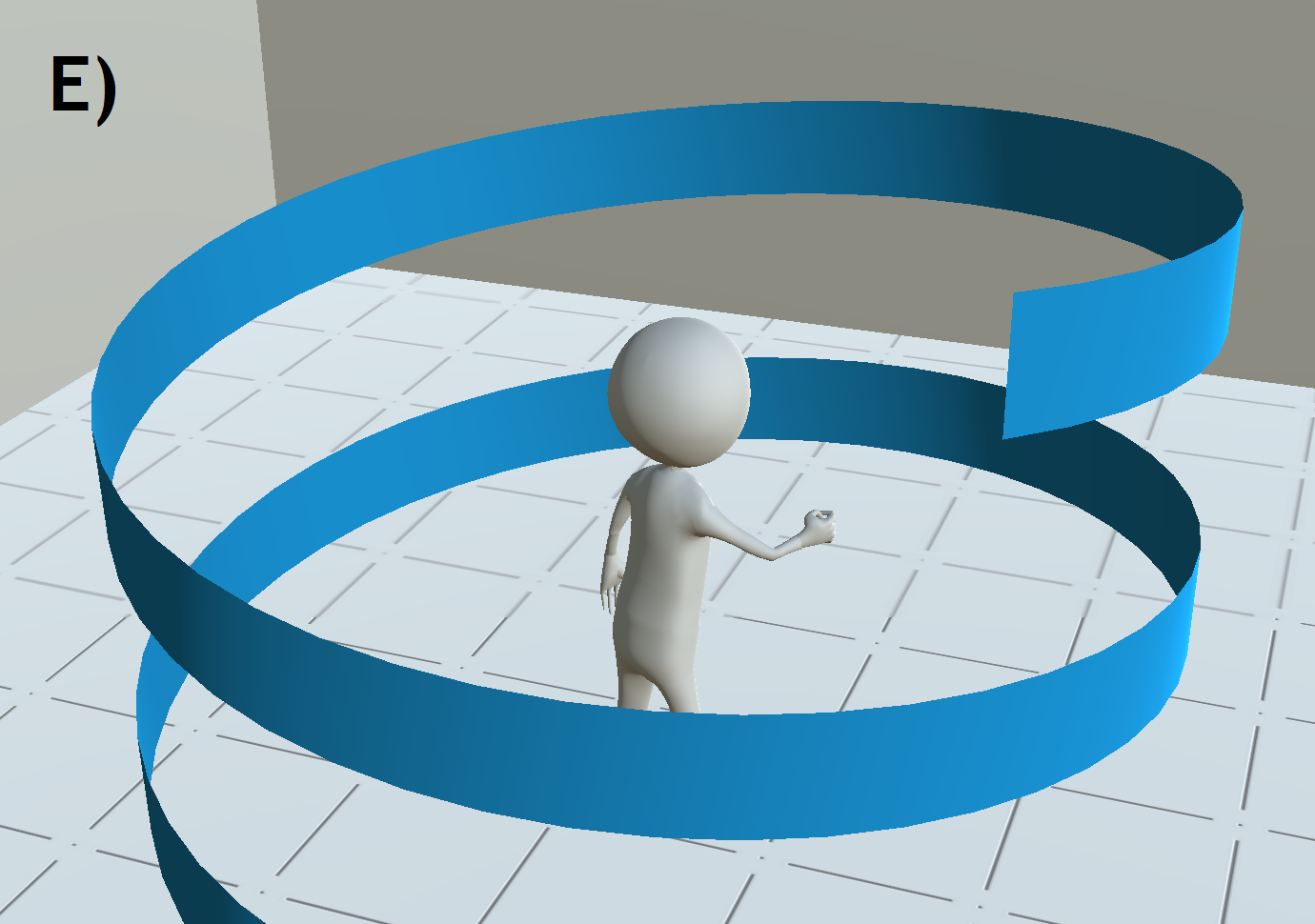} \hspace{0.05cm}
\includegraphics[width=0.23\textwidth ]{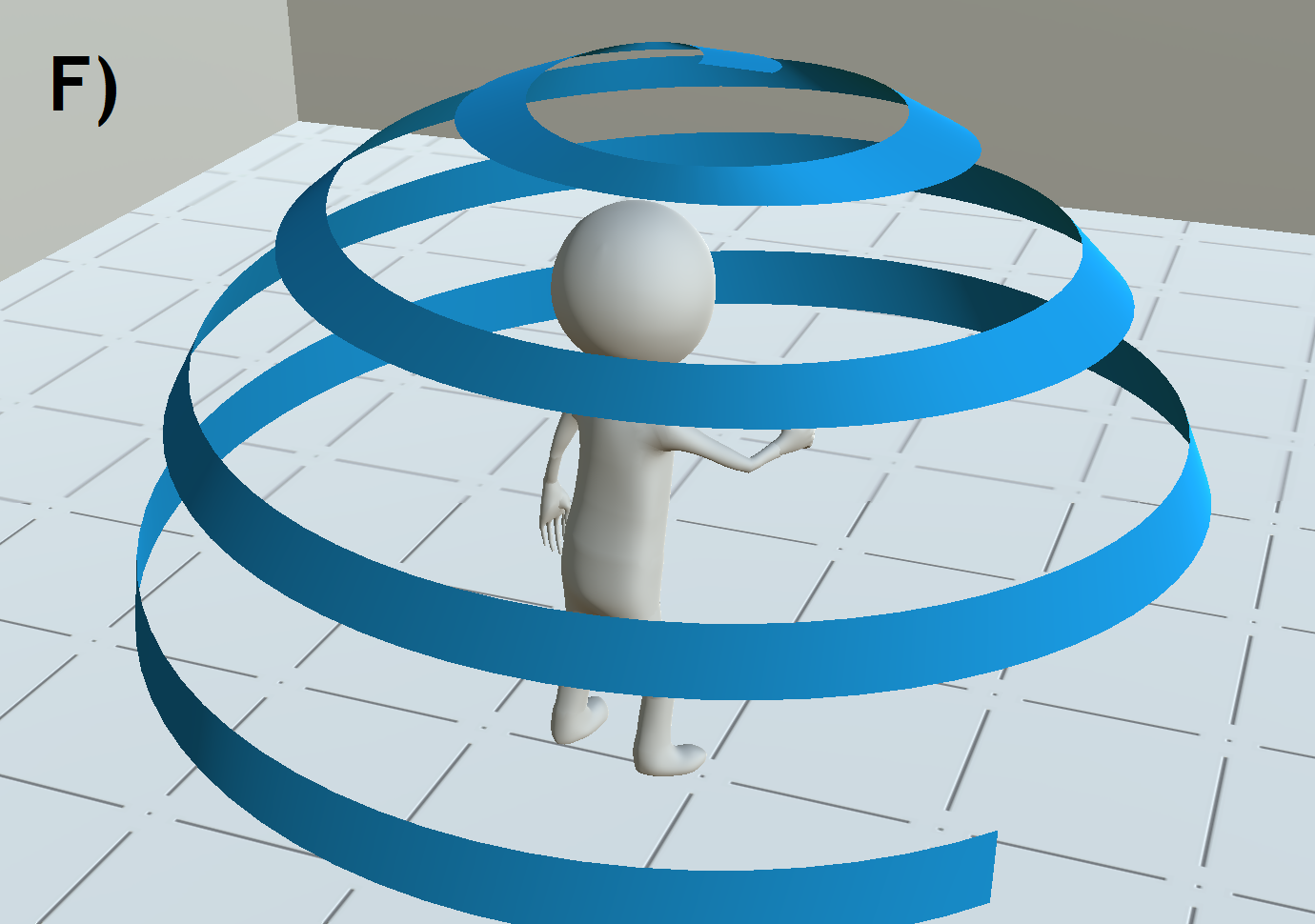} \\

\caption{Third person views of options for the representation dimension of the 3D timeline design space. These illustrations show \textbf{A)}~linear, \textbf{B)}~convex arc of circle, \textbf{C)}~convex parabola, \textbf{D)}~concave parabola, \textbf{E)}~helicoid and \textbf{F)}~spherical representations.}\label{fig:representationTable}
\end{figure}

The representation of a timeline corresponds to the shape, the guiding curve along which are placed the time points.
This curve is therefore an equivalent of a time axis.
In most cases, timelines are represented linearly~\cite{brehmer_timelines_2017}, as for history timelines. 
This type of representation maintains a single orientation, often horizontal and following reading direction, for simplicity purpose.
Other designs use radial representation, to emphasize cyclic repetitions, or spirals, often used to get compact visualizations.
Depending on the goal of the author of the timeline, some other types of representation can be designed for aesthetic, density or pedagogic purposes.
From these representations used in 2D designs, we propose representations that take advantage of the 3D environment.
First, the linear representation could be extended to a \textbf{3D curve}.
Three main possibilities emerge:

\paragraph{Flat curve.}The most direct extension of the linear representation in 2D would be to use a simple flat curve, as seen in Figure \ref{fig:representationTable}-A.
In terms of representation, there is no limit to the amount of time points displayable in the environment.
The main drawback is that the further the curve goes, the harder it is to reach or interact with the time points.

\paragraph{Convex curve.}This representation displays the time points around the user.
%
%
Several types of curve could be considered.
An arc of a circle centered on the user would place all time points at the same distance, enabling the same capacity of interaction and observation for each time point.
Such curve is shown in Figure \ref{fig:representationTable}-B.
This counters the primary issue of the flat curve representation, yet the radius of the circle will increase with the amount of time points, limiting the proximity required for interaction.
To compromise between proximity and amount of time points, the user could be brought closer from a part of the arc of circle.
Another approach would be to use a parabolic curve, as in Figure \ref{fig:representationTable}-C. 
In both those cases, a local temporal context is brought very close from the user for interaction and observation, yet the other time points remain at a reasonable distance.

\paragraph{Concave curve.}As opposed to the previous ones, a concave curve can be used to display close from the user a few time points, and sending the other ones that are further in time, even further in the environment.
This representation emphasizes a local temporal context, frees some radial space around the user and decreases the density of information in their field of view.
In this sense, the amount of time points displayed is potentially unlimited, yet harder to access.
Similarly to convex curves, concave curves could rely on parabolas, as in Figure \ref{fig:representationTable}-D, or arc of circles, depending on the amount of time points that need to be close to the user.\\

Other designs for 2D timelines rely on circular or spiral representation.
These representations are particularly adapted to provide an insightful representation of periodicity in time.
For example, a spiral layout can be used to represent several occurrences of a same periodic event, as demonstrated in Weber et al. work~\cite{weber2001visualizing}.
However, because of the varying radius of a spiral loop, the number of period has to be limited.
Nonetheless, spirals are also intrinsically space-filling, thus producing dense representations, and 
can be engaging or even playful.
Consequently, we propose two additional representations adapted to the design of a 3D timeline:

\paragraph{Helicoid.} This representation coils the time points around a cylinder base. 
It can be centered on the user, displaying every time point close from the user, as shown in Figure \ref{fig:representationTable}-E.
Similarly to the 2D spiral representation, this shape invites the display of several occurrences of periodic data, without the drawback of a varying radius between each period.
For example, manipulating the scale dimension of the design, the circular periods of the helicoid could be aligned according to the occurrence of a particular event, for comparison purpose. 

\paragraph{Spherical.} Similarly, a spherical representation, as shown in Figure \ref{fig:representationTable}-F, coils around a sphere base.
The time points are thus displayed around the user at the exact same distance, reducing distortions, in size of orientation for instance, implied by distance or perspective.
This representation shares very similar advantages and drawbacks with the 2D spiral.
It is aesthetically interesting, but the varying radius between two circular periods restrains its use for periodic data. \\

These two representations share several other characteristics.
The exploration of the timeline is no longer horizontal as in the curve representation, but rather radial and vertical, allowing fast jumps in time.
It is compact and takes full advantage of the 3D work space.
However, due to the use of the spatial upward axis, it can limit the use of faceted layouts, as it could lead to confusion between the timeline branches juxtaposed and the next circle of the helicoid.
Other representations using either different curves or volume could be used, that may be tailored for specific analysis use cases, for an aesthetic purpose or to create mnemonic designs.
Changing the orientation in either axis can also be a way to explore original representations.

\subsubsection{Support: Exploiting the 3D Environment}

In usual displays, timeline branches are most usually laid out on the display plane.
As mentioned in part \ref{refTimeline}, a few works explored displaying timelines in 3D environments.
Beedocs~\cite{Beedocs} includes 3D perspective to animate and display the timeline at a different orientation for aesthetic purpose, while Kullberg et al.~\cite{kullberg1995dynamic} even lays the timeline on the horizontal plane of a 3D environment, using the vertical plane to display the content of the time points.

As we propose to display 3D timelines in VR, we introduce an additional dimension, the \textit{support}, to refer to the 3D shape that supports the timeline branches in such 3D environment.
We describe here the main choices that will mostly influence the timeline design.
This includes shape, size, position, and also count of supports on which the timeline branches are displayed.
We propose the following choices for this dimension:

\paragraph{Vertical plane.} The branches are laid on a plane vertically in front of the user, like a classic 2D display.

\paragraph{Horizontal plane.} The branches are laid on a plane parallel to the floor.
Horizontal head movements and field of view are usually preferred over vertical ones, thus this design choice could be beneficial compared to a vertical plane support in some cases, such as for comparison at one time point of multiple branches. The main drawback is occlusion, since branches are displayed one in front of the others.

\paragraph{Multiple Planes.} As opposed to the previous design choices involving a single plane, here the branches are displayed on several ones. This can be useful to optimize application design, or separate clearly facets and segments.

\paragraph{Cubic.} The branches are laid along multiple vertical or horizontal planes. Such layouts can be useful to organize branches along 2 parameters, much like a 2 dimensional array. While very meaningful in this purpose, this design is quite dense, which can be overwhelming for a user and cause difficulties in the exploration.

\paragraph{Concentric cylinders.} The branches are displayed on multiple isoradial planes centered on the user. In terms of exploration, it is quite similar to horizontal planes, yet fills the space around the user, which can serve aesthetic purpose.\\

Each of these dimension choices are illustrated in Figure~\ref{fig:teaser}, second line, in order. 
While this dimension is unnecessary for timelines with a unified layout, these design choices can be useful for some specific use cases, for application design, notably in terms of workspace usage, or for aesthetics.
Yet, some of these designs can also involve significant occlusion issues.
Adapted navigation and application controls might be required when using these design choices.

\subsection{Interacting with 3D Timelines}\label{basicInteraction}

This section discuses the interaction methods required for the exploration and manipulation of a 3D timeline. First, we consider methods enabling the free exploration of the timeline, supporting consume and search tasks. Then, we consider tasks than allow the manipulation of the timeline, in order to modify the arrangement of the timeline. 
The discussed interactions consider that the user is using an immersive virtual reality setup (e.g. head-mounted display), in which the user can move physically to explore the virtual environment and that additional input capabilities are provided by hand-held devices (e.g. controller). For the sake of generalization, we will not discuss how the interactions can be implemented, but which parameters (e.g. distance among time points) could be modified.
%

%
%
The timeline is arranged in the environment in such a way as to have one time point, that we will call from now on \textit{central time point} of the timeline, that is the closest from the user, as shown in Figure~\ref{fig:centralTimePoint}.
Time points are identified by an indicator beneath the objects represented.

\subsubsection{Exploring the Timeline}\label{exploring}


First, the user can physically move (e.g. walk) in order to approach the time point of interest. However, the range of time points that can be explored will be dependent on the shape of the timeline and the workspace of the virtual reality setup. 
Some timeline designs covering an important part of the work space around the user, for example when using convex curves or spherical representation, will enable the direct exploration of all time points without requiring a large physical workspace.
In contrast, linear or parabolic representations can spread far away for the user, and faceted or segmented layouts might display data too high, limiting the number of time points that the user can access.
To overcome this limitation, one solution is to consider virtual navigation methods~\cite{laviola20173d}, that enable the user to virtually (without physically moving) navigate. Potential good candidates are scene-in-hand methods~\cite{ware1990using} or virtual steering~\cite{brument2020does} which will allow users to reach time points arbitrarily far away from them. 
Yet, the choice in the virtual navigation method will depend on the implementation of the timeline, the data represented and the end user profile.

Finally, another alternative to enable a full exploration of the timeline is to move the timeline along its time axis, thus changing the \textit{central time point}. This can be done continuously by scrolling the time points (e.g. using a temporal slider), or directly by selecting a time point from afar (e.g. using a ray-based selection method~\cite{argelaguet2013survey}). These methods will minimize the user's physical motion.


%
%
%

%

%

\subsubsection{Manipulating the Timeline}\label{manipulatingView}


%
There are numerous continuous parameters involved in the display of the timeline, such as the space between time points, the center and radius of curvature for curved representations, the height for helicoid representations or the distance between timeline branches for faceted and segmented layouts.
These values can strongly vary, notably depending on the amount time points displayed in the timeline.
For this reason, the user could be able to choose among pre-defined configurations of timeline designs. 
Nonetheless, to increase the flexibility of the visualization, the user should be able to adjust the parameters of the timeline display in order to better fit their use-case requirements.
The modification of these values can be done using graphical user interfaces or specific 3D widgets~\cite{dachselt2007three}.
%
Moreover, in order to control the information contained in the environment level-of-detail options are also relevant.
The user should be able to choose to display only a proportion of the time points selected regularly along the timeline, and to directly choose time zones to collapse or extend via manual selection.

\subsubsection{Manipulating the Content of the Time Points}\label{filter}


The final interaction category relies on how the content of the time points could be modified. 
First, the 3D objects at each time point can be rotated and scaled, in order to enable the user to explore them from different orientations. To ensure consistency, we advocate to link the rotations and scale for all time points, thus obtaining a consistent orientation for all time points. Such manipulation could be achieved using bi-manual manipulation methods~\cite{mendes2019survey}. 

Second, operations to reduce the information at each time point can also be considered. 
(A) Cut-away views defined by a clipping plane or a 3D volume (e.g. cube) operators. The user could define the position and orientation of the cutaway operator in the central time point, a corresponding cut-away view could be applied as well on each of the other object or group of objects, relatively to their barycenter. Standard 3D manipulation techniques can be used to control the cutaway operator~\cite{mendes2019survey}.
(B) Object and value filtering operations. These operations will be highly dependent on the available data of the S3D. For example, if the data for each time point is composed by a set of 3D objects (e.g. different 3D meshes, segmented 3D volume), individual objects could be removed from the timeline by selecting them. Moreover, if the individual objects are linked with other objects at different time points (e.g. same object appearing at different time points), the selected objects and their linked objects could also be hidden. Finally, if additional categorical, ordinal or numerical attributes are available for each object (e.g. volume of a given object), filtering operations could also be defined (e.g. hide the objects with a volume lower than a threshold). 
These operations to reduce the displayed information could be of great relevance to reduce the amount of displayed information, reducing clutter. These interactions could be designed considering existing 3D selection methods~\cite{argelaguet2013survey} and 3D graphical user interfaces~\cite{dachselt2007three}.

%
%
%




\subsection{Choosing a Timeline Design}\label{viable}

In this part, we propose an set of design criteria that will drive the design choices of 3D timelines.
Focusing on the constraints that can be introduced due to the attributes of the data (topology, resolution, rendering method), application design constraints and more importantly the narration, storytelling and expressiveness purpose, we identified the following key criteria:

\paragraph{Preferred exploration.} Designs will favor exploration of the timeline simply through head rotation, or through virtual navigation and manipulation of the timeline.

\paragraph{Limit of time points.} The amount of time points displayable objects easily discernible can be limited, depending on how the design partitions the workspace.

\paragraph{Workspace usage and visual clutter.} Designs occupy more or less radial and vertical space around the user, which can cause occlusion and application design issues.

\paragraph{Periodicity in the data.} Periodicity and temporal patterns are often important characteristics in data analysis, and some design choices, such as segmentation or helicoids can emphasize such characteristic.

\paragraph{Aesthetics, originality, playfulness.}The attractiveness of a visualization can be crucial when trying to convey information. The design can be adapted to the use case and the public for this purpose.\\


Innovative designs might create more adapted visualizations or more catchy aesthetics.
Yet, it should be noted that exotic designs might also be more difficult to interpret, as stated by Brehmer et al.~\cite{brehmer_timelines_2017}, it is thus recommended limit the unfamiliar choices in the design of the timeline.
On the other hand, in some cases design choices are interlocked together between several dimensions.
For example, a helicoid representation strongly implies the use of a cylinder support.

\section{Evaluation} \label{userExp}

The previous section discussed a wide range of potential 3D timeline representations, yet, from the design on itself, it remains unclear how users will leverage the 3D timeline to explore S4D data. Thus, we conducted a user evaluation in order to assess the potential benefits of 3D timelines for the exploration of S4D data. Our objective was to assess representative 3D timeline designs to explore how users will perform exploration tasks.

\subsection{Dataset and Task Design}

\begin{figure*}[h]
\centering
\includegraphics[width=0.95\textwidth]{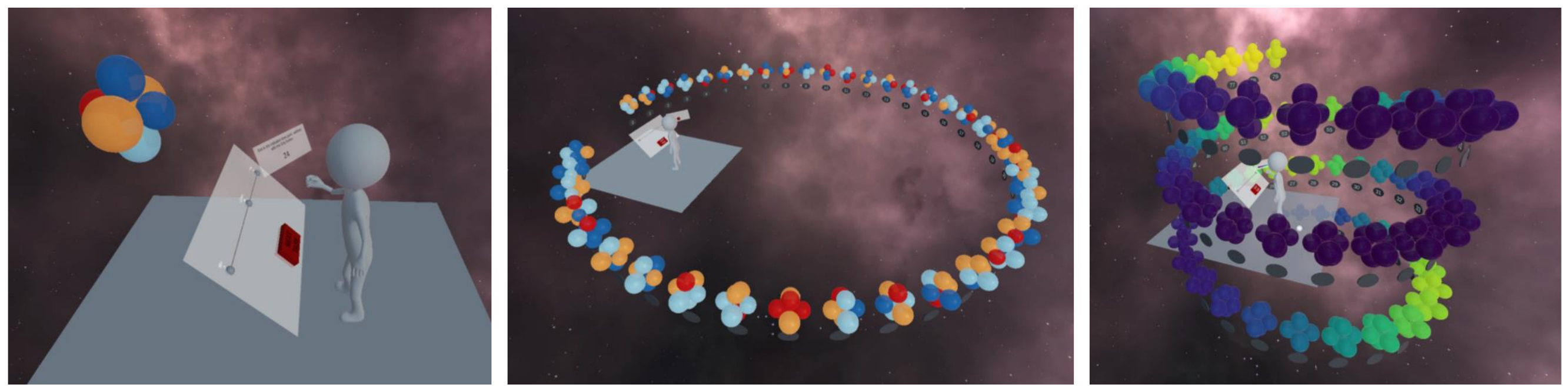}
\caption{Environment for the user experiment. The right image shows the \textbf{Locating time point} task in the \textit{No timeline} condition; the middle one shows the \textbf{Counting occurrences} task in the \textit{Curved} condition; the left one shows the \textbf{Finding a global maximum} task in the \textit{Helicoid} condition, where the objects are colored using the \textit{viridis} colormap.}\label{fig:userExperiment}
\end{figure*}

%
%

The S4D dataset used in the evaluation was generated procedurally in order to keep the experiment accessible but also to limit the potential bias between users in the approach of complex S4D and in the manipulation of 3D models (see Figure~\ref{fig:userExperiment}).
As such, we limited the number of time points and objects ensuring that users will be able to apprehend the dataset. The objects were represented by spheres, and their position did not change over time. The dataset was composed of 6 objects evolving over time, and depending on the task (see below) having either 40 or 80 time points.
The size of the objects increased monotonously, so that users could keep points of reference when exploring over time. 
In order to enrich the dataset, objects were annotated with 2 types of information.
First, a categorical information describing the affiliation of each object to one of 5 \textbf{groups}, which could change over time (see Figure~\ref{fig:userExperiment}, center), was encoded with different colors.
We introduced randomly several occurrences of a specific pattern in the group information, on which we can evaluate success and error rates for the related task.
Second, a continuous information (float \textbf{value}) evolving over time, encoded with the \textit{viridis} color map (see Figure~\ref{fig:userExperiment}, right). 
This value was generated by adding 4 random Gaussian functions, one on each 20-time-point segment of the dataset.

During the experiment, the objective was to let participants to explore the dataset and perform different tasks. We defined four data exploration tasks that corresponded to common tasks for the exploration of time-varying data: locating a time point, counting occurrences of a particular event,  finding temporal patterns or finding a global maximum. 
For the \textbf{location task}, a dataset of 80 time points was used. Participants had to find random defined time points between 0 and 79. They had to reach the indicated time point by using the time exploration technique available, described below.
For the \textbf{counting task}, a dataset of 40 time points, annotated with the group information was used. Participants had to count the number of objects of a given group in all time points. To limit the length of the task, it was limited to 3 minutes.
For the \textbf{pattern task}, a dataset of 40 time points, annotated with the group information was used. Participants had to count the appearances of a specific 3-time-point-long temporal pattern that objects might follow in the 40 time points displayed. The task was limited to 3 minutes of exploration.
Finally, for the \textbf{maximum task}, a dataset of 80 time points, annotated with the continuous information was used. Participants must find the time point exhibiting the maximum value. For simplicity, all objects from the same time point had the same value. 





\subsection{Hypotheses}

The two first tasks (location and counting) did not require temporal context. 
Indeed, the information to find was punctual in time, thus any method allowing scrolling through time could be sufficient.
Thus, we did not expect timelines to be particularly efficient for this type of task.
Although having too much information can be overwhelming for the user, we argue that the added temporal context could still be beneficial.
On the other hand, the two other tasks (pattern and maximum) required information from previous or following time points.
As such, we expected better outcomes using timelines.
The result could also be strongly dependent on the timeline design, as how the information is displayed and accessed.
%
We expect that denser timeline designs would allow to access larger temporal context and could thus perform well when it is required, while also being useful to handle higher amounts of time points.
Accordingly, we chose different timeline designs, which are detailed in Section~\ref{sec:Variables}.
In summary, our hypotheses were:

%
%
%

\begin{description}
\item[H1a]Tasks that did not require exploring the temporal context will be achieved at comparable performance with or without timeline.
\item[H1b]Tasks that did not require exploring the temporal context will be achieved more comfortably with a timeline.
\item[H2]Tasks that require exploring the temporal context will be achieved better using a timeline.
\item[H3]Denser timeline designs will more adequate for achieving the tasks.

\end{description}


\subsection{Apparatus and Participants}

The material used for the tests and experiences described throughout this paper was as follow: a HTC Vive Pro HMD with 2 Vive controllers, a PC with Windows 10, an Intel Xeon W-2104 CPU (4 cores, 3.2 Ghz base frequency) and a Nvidia RTX 2080 GPU. The VR application runs on Unity 2019.4.5f1.

This experiment included 21 participants, 5 women and 16 men, aged from 22 to 42, mostly recruited from the local laboratory.
Two of them had low experience with VR, the others were expert VR users. 

\subsection{Independent and Dependent Variables}\label{sec:Variables}

The experiment followed a within-subject design in which participants had to perform the different tasks using 3 different visual conditions. The first one was the \textit{No timeline} condition, in which only a snapshot of the dataset at an instant was displayed. The two others used timeline designs which took advantage of the 3D workspance, a helicoid unified and a circular convex unified timelines. They will be referred to as respectively \textit{Helicoid} and \textit{Curved} conditions.

For all tasks, we measured the time to complete the task, as well as the movements and the time using the different available VR interaction tools. 
For the locate task, we also computed the difference between the selected time point with the target time point.
For the counting task, we also computed the error rate. 
For the pattern task, we further evaluated the precision and recall.
Finally, for the maximum tasks, we evaluated the accuracy.

\begin{figure*}[t]
\centering
\subfloat[][Comfort]{\includegraphics[width=0.45\textwidth]{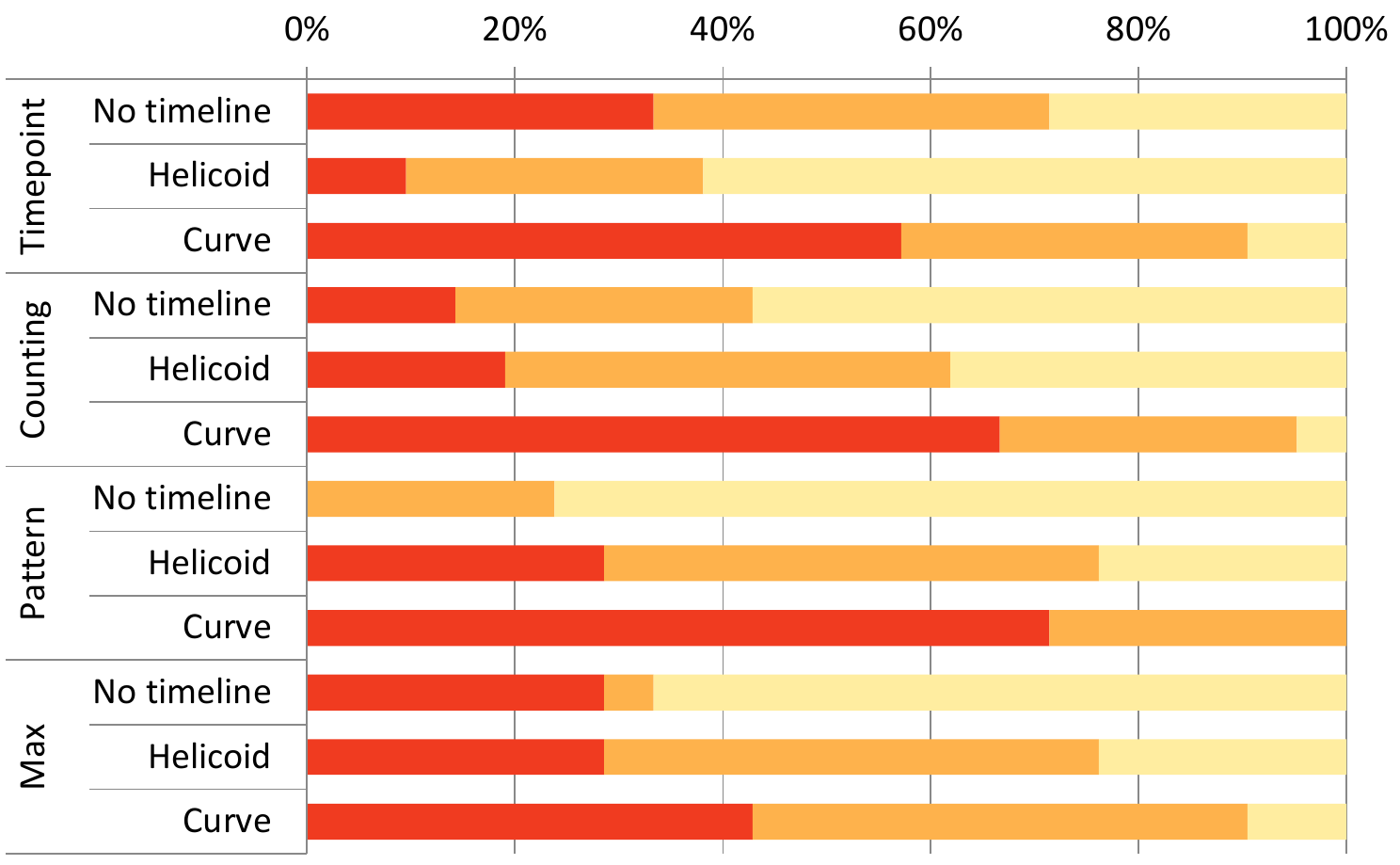}} \hspace{0.5cm}
\subfloat[][Efficiency]{\includegraphics[width=0.45\textwidth]{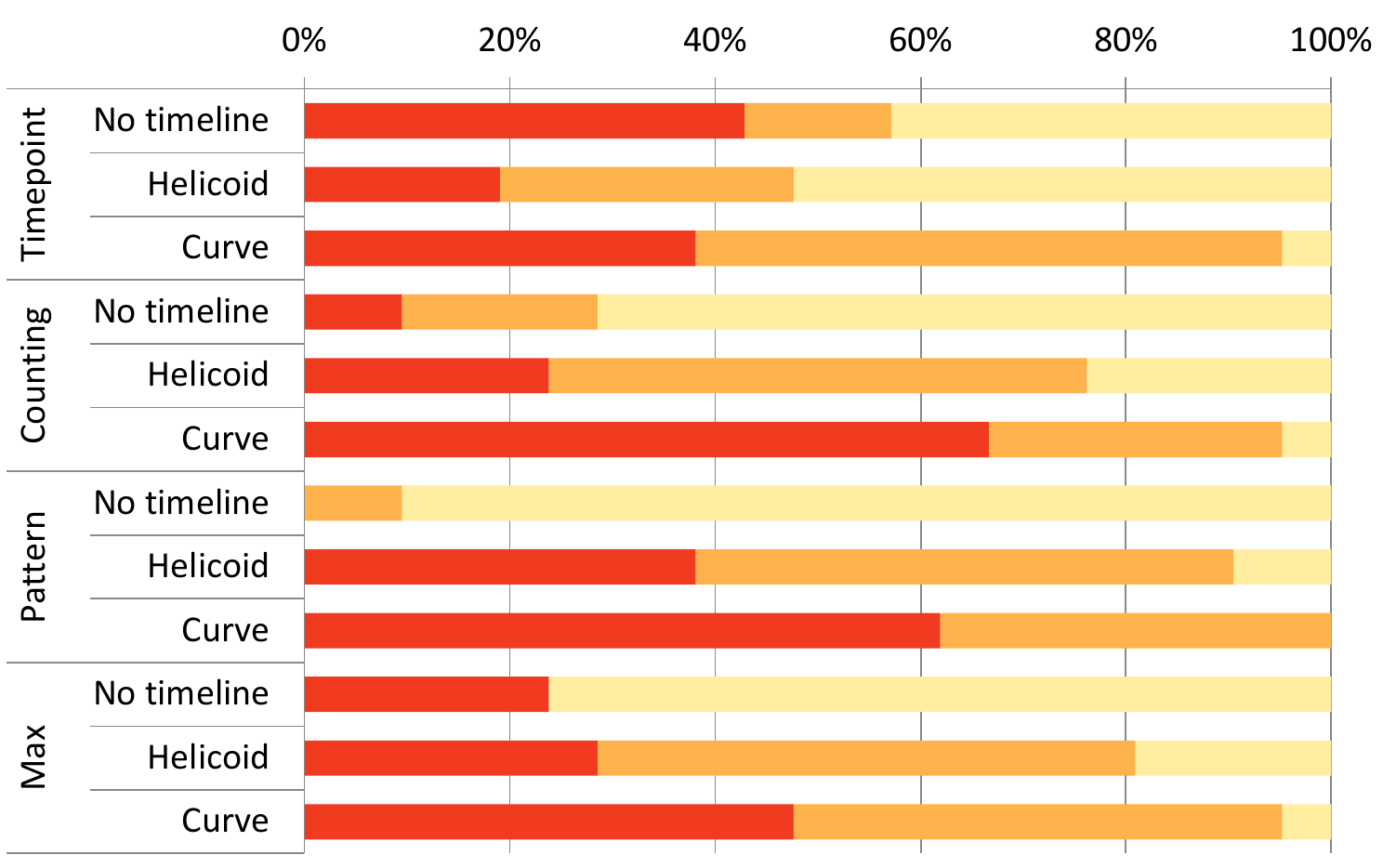}}\\

\includegraphics[width=0.3\textwidth]{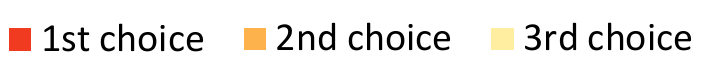}
\caption{Subjective results for each timeline design and task. (a) Ranking results for the question ``What condition felt the most comfortable for this task?''. (b) Ranking results for the question ``What condition seemed the most efficient for this task?'' }\label{fig:subjective_results}
\end{figure*}

\subsection{Experimental Protocol}

At the beginning of the experiment, participants signed a consent form and filled a demographic information questionnaire.
Participants received an explanation of the VR system and the available interactions. We mainly considered interactions that allowed the users to achieve the designed tasks. 
%
%
As such, object selection and task validation was done using a virtual laser pointer, and object manipulation was done using a bi-manual control.
Time exploration in the \textit{No timeline} condition was done by scrolling on the controller touchpad or using a direct manipulation on a slider (see Figure~\ref{fig:userExperiment}, left).
In the other conditions, scrolling made the timeline move along its directing curve, and a direct selection based on a gaze and click interaction was available, users had to orient their head towards the time point and press a controller button.
We let them get used to the tools and application layout for 10 minutes.
The participant then proceeds to the tasks in order, in which the display condition was counterbalanced using a Latin Square.


After each condition, participants were asked to fill the  SUS~\cite{brooke1996sus} questionnaire to evaluate usability, and the
%
CSQ-VR~\cite{kourtesis2019validation} and FMSS~\cite{keshavarz2011validating} questionnaires to assess cybersickness.
Finally, to subjectively assess the timelines, for each task and condition we asked the following questions: ``What condition felt the most comfortable for this task?'' and ``What condition seemed the most efficient for this task?''.

\subsection{Results} 

For each task, we obtained a total of 63 observations, characterized by the 3-level between-subjects factor \textbf{Order} and 3-level within-subjects factor \textbf{Condition}.
We used repeated measures two-way ANOVA to analyze the results.
When the normality assumption was violated (Shapiro-Wilk normality test) we transform the data using the Aligned Rank Transform (ART)~\cite{wobbrock2011aligned} prior conducting the ANOVA analysis.
Post-hoc pairwise tests with Bonferroni correction ($\alpha=0.05$) were used when needed.


\subsubsection{Main Effects for Condition}

\textbf{Locate task.}
This task was the first of each condition, thus we removed the two first trials of each participant, which might have induced too much variability, as users were still getting used to the method, and averaged the 8 other observations.
The users made almost no errors (\msd{0.018}{0.077}), showing that participants were able to perform the task accurately.
In addition, the ART ANOVA order and condition vs \textbf{average time} showed a significant effect of the condition (\ARTanovatest{2,36}{13.87}{0.001}). Post-hoc tests validated lower times during the \textit{No timeline} condition (\msd{7.12}{1.86}) compared to the two others (\textit{Curved} \msdp{8.75}{2.38}{0.001}, \textit{Helicoid} \msdp{9.57}{4.13}{0.001}), yet no significant difference between the latter.
This tend to \textbf{invalidate H1a}, since performance was lower using timelines.
%
 \vspace{\itemsep}



\textbf{Counting Task.} While the task was limited in time, the limit was never reached. 
The ART ANOVA order and condition vs \textbf{time} showed a significant effect of the condition (\ARTanovatest{2,36}{11.99}{0.001}).
Post-hoc tests validated that the \textit{No timeline} condition (\msd{106.8}{40.4}) gave significantly higher times than the two others (\textit{Curved} \msdp{76.8}{42.2}{0.01}, \textit{Helicoid} \msdp{68.4}{32.7}{0.001}), yet no significant difference between the latter.
%
%
This tends to \textbf{invalidate H1a}, since performance was higher with timelines.
The ART ANOVA order and condition vs error rate gave no significant results. \vspace{\itemsep}



\textbf{Pattern task.} Participants were able to achieve the task with almost no errors showing that all methods allowed to perform the task (\msd{0.993}{0.025}).
Moreover, the ANOVA order and condition vs \textbf{recall} showed a significant condition effect (\anovatest{1.63,29.36}{6.23}{0.01}{0.26}).
Post-hoc tests validated a significantly lower recall between the \textit{No timeline} condition (\msd{0.652}{.201}) and the two others (\textit{Curved} \msdp{0.798}{0.15}{0.05}, \textit{Helicoid} \msdp{0.818}{0.17}{0.01}), yet no significant difference between the latter was found.
This tends to \textbf{support H2}.

The ART ANOVA order and condition vs \textbf{time} showed a significant main effect on condition (\ARTanovatest{2,36}{12.5}{0.001}).
Post-hoc tests showed significantly lower times for the \textit{Helicoid} condition (\msd{157}{38.8}) against the \textit{No timeline} condition (\msdp{192}{20.9}{0.001}), and also against the \textit{Curved} condition (\msdp{180}{28.5}{0.05}), yet no significant difference between the latter.
This tends to \textbf{support H3}.
 \vspace{\itemsep}



\textbf{Maximum task.} We averaged the measured values of each of the 5 trials of each participant on each condition.
The measured error rate was always close to 1 ($M=0.997$, Min=$0.98$), participants were able to accurately perform the task with all conditions.
Furthermore, the ANOVA order and condition vs \textbf{average time} showed a significant effect of the condition (\anovatest{1.69,30.43}{8.27}{0.01}{0.31}).
Post-hoc validated significantly higher times for the \textit{No timeline} condition (\msd{32.5}{10.54}) than the \textit{Helicoid} condition (\msdp{26.1}{9.12}{0.05}) and the \textit{Curved} condition (\msdp{22.8}{9.19}{0.001}).
This tends to \textbf{support H2}. \vspace{\itemsep}



\textbf{Questionnaires.}
According to the CSQ-VR and FMSS results (respectively \msd{0.76/36}{0.85} and \msd{1.7/20}{1.5}), participants did not suffer a significant increase on cybersickness \ferran{provide global scores, to see if it was low}.
The ANOVA order and condition for the SUS scores showed a significant effect of condition ($p<0.05$). Post-hoc tests only showed a significant effect for the \textit{Curved} condition (\msd{81.9}{11.7}) against the \textit{No timeline} condition (\msdp{70.4}{12.2}{0.01}).


Figure~\ref{fig:subjective_results} provides a summary for comfort and efficiency assessment for each condition.
%
%
In terms of comfort, the users generally preferred the \textit{Curved} condition for every task.
The \textit{No timeline} condition was second only in the Timepoint task, for the other tasks the \textit{Helicoid} condition was preferred.
About subjective estimation of efficiency, the \textit{Curved} condition was still preferred in all tasks, except in the Finding Timepoint task, for which the \textit{No timeline} condition was arguably even.
However, the second place was more clearly attributed to the \textit{Helicoid} condition for this criteria.


\subsubsection{Order and Interaction Effects}

We analyzed potential order and interaction effects for each test.
For most, the order effect was not significant, or could not be confirmed by post-hoc tests.
However, we did find significant effects for some cases.

The ART ANOVA order and condition vs \textbf{average time} on the \textbf{Finding Timepoint} task showed a significant order effect (\ARTanovatest{2,18}{5.00}{0.05}), which was validated by a post-hoc test as a difference when the \textit{Helicoid} was the first condition (\msd{10.05}{4.26}) and when it was the last (\msd{7.1}{1.24}), $p<0.05$. 
Judging from the important standard deviation, we hypothesise that starting with the \textit{Helicoid} might have been overwhelming to some users during their familiarization with the system, causing higher achieving times in this first task of the experiment.
This interpretation is further supported as the \textit{Helicoid} was judged the least comfortable condition for this task.

The ANOVA order and condition vs \textbf{average time} on the \textbf{maximum} task showed a significant interaction effect (\anovatest{1.69,30.43}{8.27}{0.01}{0.31}).
Post-hoc tests confirmed that the results from the \textit{No timeline} condition as first condition of a participant were significantly lower than any other case when the participant executes the task in the second timeline condition.
This can be explained by a learning effect on the use of timelines, which becomes significant when achieving the task with a second design.
Though, the condition effect showed that both \textit{Curved} and \textit{Helicoid} conditions obtained significantly lower times regardless of order, so this did not alter our conclusion on H2.



\subsection{Discussion}

The results from the two first tasks, that did not essentially require temporal context, were mitigated, so we must \textbf{reject H1a}, \ferran{however the sign is different, for the first task, the no timeline was better, while for the second was the opposite, learning?}\gwendal{Well, there was an order effect... idk}.
The results from the questionnaire however \textbf{support H1b}, as users did report preference for timelines for these tasks.
\ferran{Revise this sentence $->$}Several results from the tasks of finding patterns and global maximum \textbf{support H2, that we can accept.}
Finally, results from the task of finding patterns \textbf{supports H3}.

Overall, these results showed significant benefits of using timeline designs over classic visualization for temporal exploration tasks, in terms of time and quality of completion, yet also in user experience.
Nonetheless, we expected a more significant difference between the two timeline designs, notably regarding the different number of time points between the tasks.
We explain this lack of differences due to the choice of using a Latin Square ordering method: in two out of three orderings, the \textit{Curved} condition was tested after the \textit{Helicoid}. 
As learning effects did disturb some results, we think that extending this study with the 6 other ordering possibilities might make such effect appear clearly, but would however require a large number of participants.

\section{3D Timelines to Explore 3D Temporal Imagery}

In order to illustrate the 3D timeline usage in a real application, we considered spatio-temporal datasets acquired through live embryo microscopy imaging.
Our goal was to gather domain users feedback, on how 3D timelines could be used for the exploration of real datasets. 
A total of 20 biology experts participated, among which 5 were familiar with the dataset or experts in embryology, and the others are specialized in other domains, which allowed us to gather their feedback on the usage of 3D timelines for data analytic purposes. 
They all had low to no experience with VR.

%
%


\subsection{Presentation of the Use Cases}\label{datasets}


The dataset we focused on (see Figure~\ref{fig:centralTimePoint}) is a live recording of the embryonic development of a Phallusia Mammillata, a tunicate of the ascidian class.
The ascidian embryos are characterized by their fast development and low number of cells (a few hundreds).
These species presents particularly transparent membranes, a lack of cell migration and apoptosis, i.e. programmed cell death, in its early development, simplifying the imaging process and the tracking computation.
The acquisition was made using confocal multiview light-sheet microscopy~\cite{guignard_contact-dependent_2018}. 
It generated 180 3D images, taken once every 89 seconds.
The output data was then segmented and the object tracking, or cell lineage, was computed using the ASTEC pipeline~\cite{guignard_contact-dependent_2018}, resulting in a surface-based spatio-temporal dataset of a few hundred megabytes.
Additional categorical and numerical data were added through automatic and manual process by the community of biologist working with these data.
More information on the dataset is available on MorphoNet website~\cite{leggio_morphonet_2019}.


The visualization of the post-processed dataset is an important part of the analytical process.
Initially, as most of the image processing and annotations are automatic, users need to be able to validate, correct and add data manually, for instance to check for errors in segmentation which could lead to abnormal cell shapes.
Then, the 3D nature of the data and high temporal resolution allows a general observation of the main dynamics involved in biological processes through different angles, giving insight and understanding of the evolution of the embryo.
Specific regions of interest can be identified through the exploration of the cells, their surrounding and their evolution in time. 
Thus, information such as outliers or asymmetrical behaviours between sister cells can be extracted, as shown in Figure~\ref{fig:fission}, to be ultimately compared either with a ground truth, a model, or other samples potentially acquired in different setups.

\subsection{Application and Timeline Design}

\begin{figure}[t]
\centering
\includegraphics[width=0.45\textwidth]{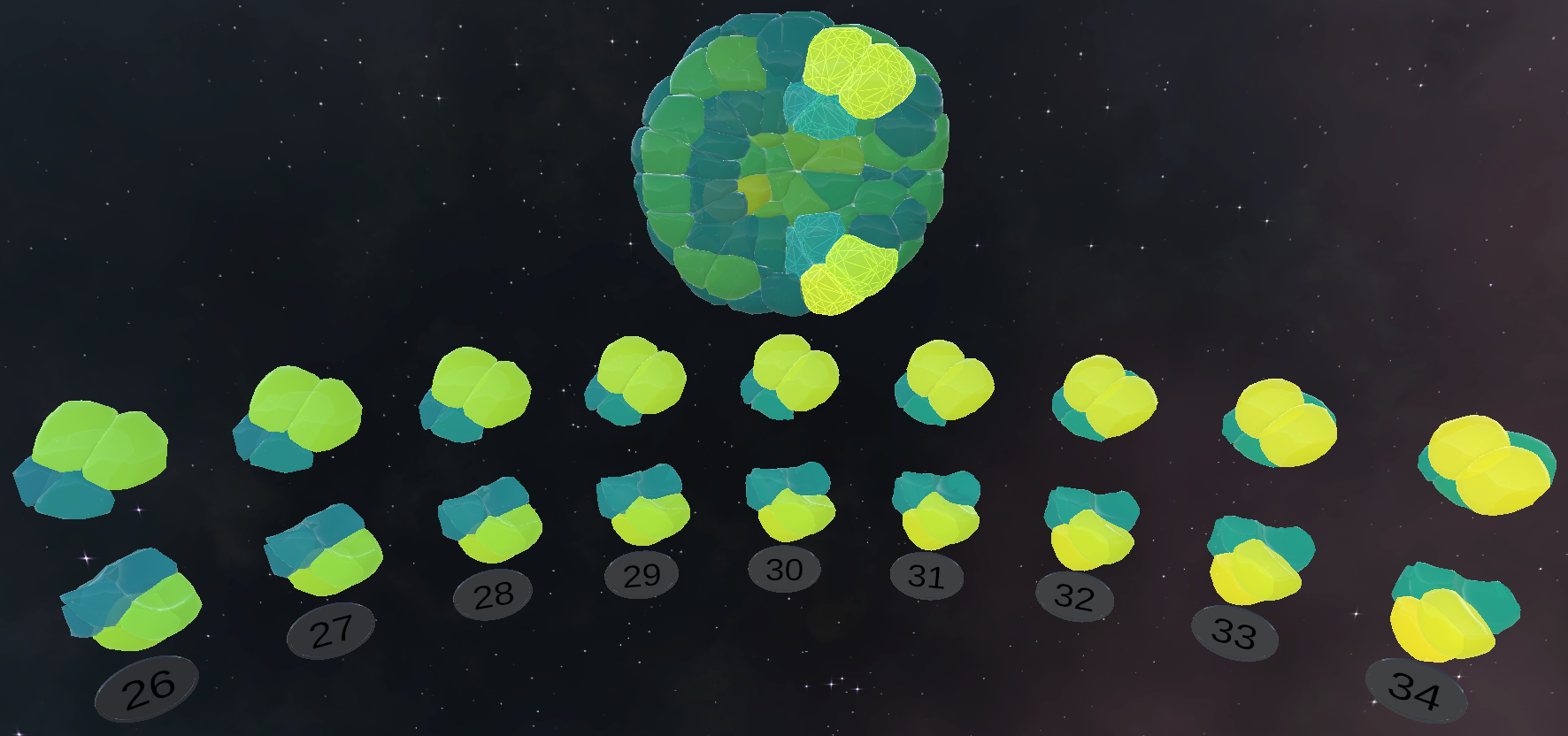}
\caption{Embryo at the 30th time point. The snapshot at the central time point shows 2 groups of are selected symmetrically as to compare their individual evolution. They are displayed on a \textbf{Circular Convex Faceted} timeline, \textbf{linear scale}, \textbf{vertical plane}. The information color-mapped is the remaining lifespan before division (colormap \textit{viridis}). }\label{fig:centralTimePoint}
\end{figure}

%
The application used in the experiment was extended in order to better support the analysis of the S4D dataset.
First, the user could choose the annotated information displayed through a 3D GUI. Their choice will change the color-mapped information. 
Second, we also implemented an object filter operation to decrease the amount of information displayed on the timeline. User could select a cell on the 3D snapshot to select the whole \textbf{4D object}, i.e. each occurrence of the cell (including parent and child cells) at each time point. Only the selected 4D objects were displayed (see Figure~\ref{fig:centralTimePoint}).

%
%
%
%

From the dataset and the different use case analyzed, we proposed the following timeline design.
%
As the dataset observed comes from imaging taken at regular time step, the scale dimension of our design is thus intrinsically \textbf{chronological and sequential}.
Considering the tasks described in the previous section, we considered two layout choices, one that could allow comparison between cells, and another for an analysis of the spatial context, respectively \textbf{faceted} and \textbf{unified} layout. 
The choice in representation is influenced by the amount of time points and the size at which the objects have to be displayed for a proper visualization.
A helicoid representation could handle the amount of time points, yet it is not optimal to use with a faceted layout if the user wants to compare many cells, in which case we preferred a curved representation.
Finally, we end up with two 3D timeline designs: (1) helicoid unified, chronological and sequential, cylinder support and (2)  curved faceted, chronological and sequential, vertical plane support.

Finally, taking advantage of the large work space offered by immersive environment, we juxtaposed the 3D timeline with the 3D snapshot at an instant or interest, as shown in Figure~\ref{fig:centralTimePoint}.
The timeline provides with temporal context, and the 3D snapshot provides with spatial context, giving a more complete point of view to the user.
In the following, we will describe examples of processes for analysis relying on the timeline visualization. The accompanying video provides additional insights on the different examples. 


%

%

\subsection{Finding Regions of Interests in Large Datasets}

Defining a region of interest is a classic visualization task, especially difficult when exploring large multidimensional datasets.
On this use case, region of interests were defined both in the 3D space and in time.
We proposed an iterative interaction process to define and refine them, both in time and in space, using 3D timelines in VR.

Initially, the user selects a few objects via ray-based selection as well as an adapted layout for the 3D timeline.
The first step of the exploration consists in a broad overview of the objects and color-mapped values displayed over the timeline, supported by the exploration techniques detailed in Section~\ref{exploring}.
During this step, the 3D snapshot will follow the gaze of the user, in order to keep spatial context while exploring the global temporal context.
Such global exploration can help eliminating some features, through value filtering or by collapsing the timeline at places nothing of interest happens, as described in Section~\ref{filter}, and identify some time points of interest.
In the second step of this process, a time point can be brought in front of the user easily through a gaze and click interaction.
The user is thus near a selected local temporal context, visible on the timeline, and the 3D snapshot is displayed in front of the user, allowing the refinement of the object selection, i.e. of the local spatial context.
Further refinement of the spatio-temporal context can be done by repeating this process.

Overall, the timeline should give enough temporal context to find times of interest, and the 3D snapshot enough spatial context to find regions of interest.
The juxtaposition and synchronisation of interaction of the two visualizations, the timeline and the 3D snapshot, allows the search and definition of a spatio-temporal region of interest. 

\subsection{Comparing Objects}

Comparison is also a usual analysis task in visualization, yet Kim et al.~\cite{kim_comparison_2017} report that this topic is quite underexplored in the case of S4D datasets. 
Timeline visualizations intrinsically juxtapose and allow a comparison between the selected objects or regions of interest at various time points.
Depending on the temporal distance, such comparison can be supported by adequate 3D timeline designs, especially in the case of periodic data, as discussed in Section~\ref{viable}.
Otherwise, interactive level-of-detail and timeline collapse methods described in Section~\ref{manipulatingView} allow to choose specific time points to compare.

However, the task of comparing 4D objects is more complex, requiring to compare both temporal and spatial features that need to be identifiable in each of the compared objects.
%
%
For example, using a faceted layout, the 3D timeline visualization allows the comparison of selected 4D objects or regions of interest, extracted from one or more datasets.
The user can create a side-by-side view of the objects of interest, and manipulate them in coordination, according to methods described in Section~\ref{filter}.
It enables a direct comparison of the 4D objects through several timeline branches, allowing to compare both the spatial information at each time point, and the evolution over time of the objects.

\subsection{Fusion and Fission}\label{fusionFission}

\begin{figure}[t]
\centering
\includegraphics[width=0.45\textwidth]{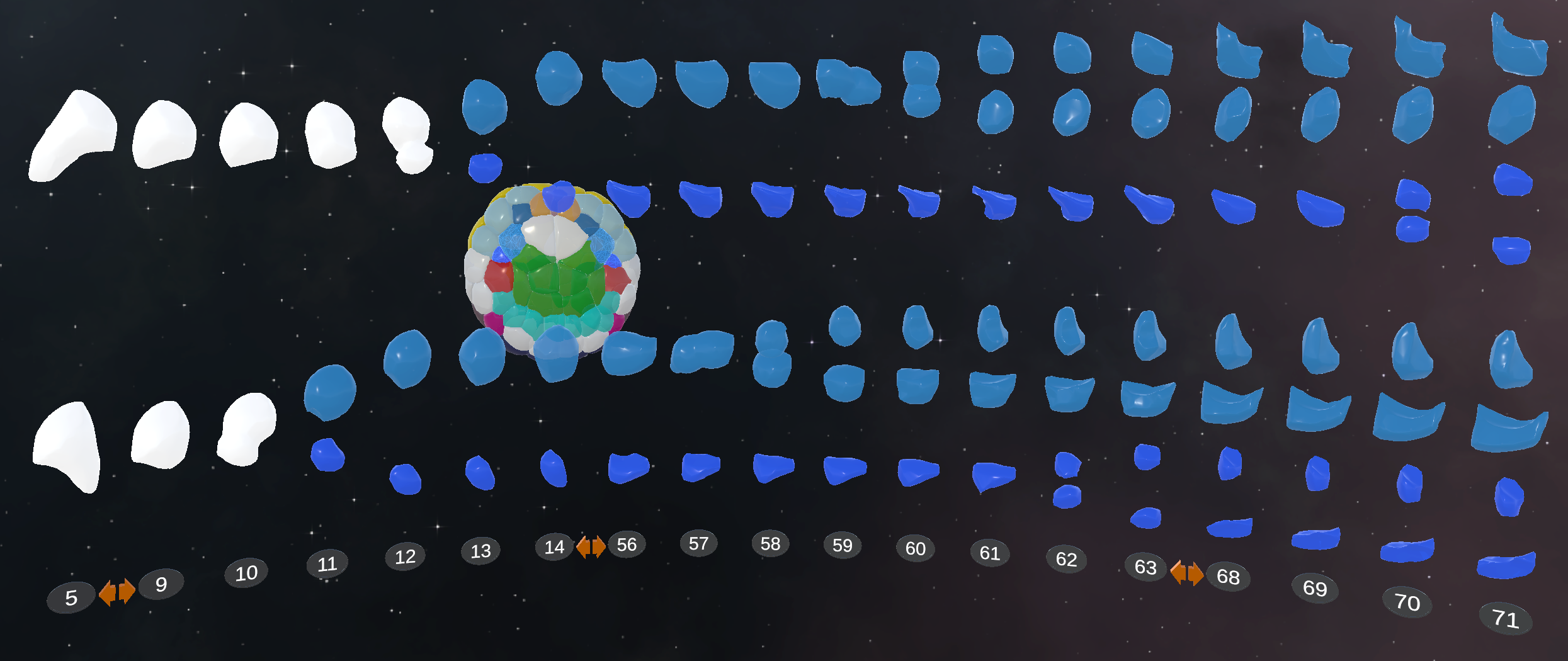}
\caption{Two sister cells are selected and displayed on the timeline. The information color-mapped corresponds to the fate of the cells. The selected ones start without annotation and are determined after a first division. The tree-like structure of the timeline shows the asymmetry of the division, with one cell significantly larger than the other.}\label{fig:fission}
\end{figure}

A particularity of the considered dataset is the notion of lifespan, as cell fusion and fission events will modify the number of 3D objects in a given time point. 
We approached this issue considering that selected 4D objects comprise all of the 3D objects involved in the event, creating different branches.
In case of a fusion event, the object will have two branches before and one after the event; in case of a fission, there will be one branch before and two after the event.
We propose an adapted layout for the 3D timeline design to represent the branches.
The resulting layout is a graph of temporal relation between the objects in the dataset.
In our specific use case, as there is mostly fission events, i.e. cell divisions, we end up with a tree visualization, oriented along a timeline, as shown in Figure~\ref{fig:fission}.
This visualization is similar to a lineage tree~\cite{sulston1983embryonic} displaying and allowing comparison of the cell shape evolution, which was one of the main tasks detailed for the use case.


\subsection{Testing and Expert Feedback}

In this section, we gather and summarize the feedback obtained during the evaluation sessions. 
They participated in sessions of about an hour by groups of 2 or 3, in order to try the application and discuss potential application of 3D timeline visualizations in their respective fields.
We presented them the dataset used and the application controls, and let them explore the data using the timeline designs proposed and the analytic techniques described previously.

The embryology specialists were overall positive about the visualization. 
They could analyse easily value evolution over different cells, as in Figure~\ref{fig:teaser}, identify outlier behaviors, and obtain information about a cell differentiation, as shown in Figure~\ref{fig:fission}, where an undetermined cell is selected and differentiate itself after division.
They also considered how such visualization could be beneficial in comparing multiple embryos, which they can difficultly do with their current visualization tools.
The specialists from other domains could not go into such details with this dataset, yet they projected how they could benefit from 3D timelines for their own data.
They quoted several examples of adapted use case with other S4D datasets such as imaging of organoids~\cite{drost2018organoids}, in context of immunology or cancer research.
They also mentioned how 3D timelines could be helpful to visualize colocalization over time of objects of different nature in multi-channel imaging.

Most of the drawbacks reported were about improving the application design to integrate it in the analysis workflow.
Several key points in the analysis process should be included to answer completely the use case, including raw data visualization, standard data formatting or annotation options.
The integration of VR equipment is also an obstacle, yet several people mentioned they would be willing integrate it in their work stations if the mentioned options were implemented.

\section{General Discussion and Conclusion}

In this paper, we proposed an extension of the 2D timeline design space proposed by Brehmer et al.~\cite{brehmer_timelines_2017} into a 3D timeline design space, focusing on their use for S4D dataset visualization.
To do so, we extended the \textit{representation} dimension, using 3D curves as guiding lines for timelines, and introduced an additional dimension describing the 3D geometry on which several timeline branches could be displayed, named \textit{support} dimension.
We proposed to use these timelines in a VR environment, leveraging the benefits of the material to enhance the exploration and interaction with such structures.
We tested two 3D timeline designs against a baseline visualization based on a 3D render and time slider, on tasks oriented toward the exploration of a S4D dataset.
The experiment results led us to conclude that 3D timelines significantly improved the achievements of tasks requiring large temporal context, yet could not conclude on the benefits of one or the other timeline.
Finally, an application implementing 3D timelines and adequate interaction methods on a embryo S4D dataset was tested by 20 biologists. 
Their feedback was very positive, both for this specific dataset but also opening to various use cases in other biology domains.

%
As opposed to the design space proposed by Brehmer, we are not classifying existing examples of timelines. 
By extending the 2D design space, we could propose a mostly usable and general 3D design space, yet it also implies our design space might be less exhaustive.
We expect that both the representation and support dimension could be even further explored, as they exploit the 3-dimension space, and could be used for more innovative or use-case specific design choices in each of these design choices.
Nonetheless, S4D data are extremely varied, and the different topology, resolution, temporal characteristics might be obstacles in the encoding of such data as time point information.
Yet, once this obstacle overtaken, this variety in the data and in the associated analysis use-cases could inspire other design choices, which would also enrich our design space.

Similarly, future works could improve and enrich the interaction pallet we provided.
The use of 6-degree-of-freedom 3D interface allowed us to implement efficient yet basic exploration and manipulation of S4D datasets for general tasks. 
However, the literature proposes finer techniques, notably for selection and navigation~\cite{laviola20173d}, that could be more adapted depending on the characteristics of the data or the 3D timeline design choices.
Our interaction pallet allows to approach most of the Immersive Analytics tasks described by Fonnet and Prié~\cite{fonnet_survey_2019}, yet some of them were left out.
Notably, the biologist who tested the application regretted the lack of annotation functionalities, as such task is key in the analysis process.
We suggest that such limitation, as well as the obstacle that is the integration of VR material in analysts workspace, could be overcome through asymmetric collaborative visualization systems, as described by Reski et al~\cite{reski2022empirical}, with one user on a desktop application and the other in VR.
